\documentclass[aps,prb,twocolumn,showpacs,superscriptaddress,10pt]{revtex4-1} 

\usepackage{graphicx}
\usepackage[utf8]{inputenc}
\usepackage{parskip}
\usepackage{xcolor}
\usepackage{amssymb}
\usepackage{amsmath}
\usepackage{mathtools}
\usepackage{physics}
\usepackage{bm} 
\usepackage[percent]{overpic}
\usepackage{microtype}
\usepackage[caption=false]{subfig}
\usepackage{makecell}
\usepackage{placeins}
\usepackage{soul}
\usepackage[pdftex]{hyperref}

\newcommand{\ie}{\emph{i.e.}}
\newcommand{\Ie}{\emph{I.e.}}

\usepackage[]{nth} 

\newcommand\unit[1]{\,\mathrm{#1}}
\newcommand\vek[1]{\boldsymbol{\mathbf{#1}}}

\renewcommand{\vr}{\vek{r}}

\newcommand{\teps}{\tilde\epsilon}

\newcommand\pa[1]{\left ( #1 \right )}
\renewcommand\pb[1]{\left [ #1 \right ]}

\newcommand\Intop[3]{\int_{\mathrlap{#1}}^{\mathrlap{#2}}\!\!\mathrm{d}{#3}\,} 
\newcommand\Intopnl[1]{\int\!\!\mathrm{d}{#1}\,} 

\begin{document}
\title{Emergent scale invariance of  non-classical plasmons in graphene nanoribbons}

\author{K\aa re Obel Wedel}
\affiliation{Department of Photonics Engineering, Technical University of Denmark, {\O}rsteds Plads, Bldg.~345A, DK-2800 Kongens Lyngby, Denmark}
\affiliation{Department of Physics, Technical University of Denmark, Fysikvej, Bldg.~307, DK-2800 Kongens Lyngby, Denmark}
\affiliation{Center for Nanostructured Graphene (CNG), Technical University of Denmark, {\O}rsteds Plads, Bldg.~345C, DK-2800 Kongens Lyngby, Denmark}
\author{N. Asger Mortensen}
\affiliation{Center for Nano Optics, University of Southern Denmark, Campusvej 55, DK-5230 Odense M, Denmark}
\affiliation{Danish Institute for Advanced Study, University of Southern Denmark, Campusvej 55, DK-5230 Odense M, Denmark}
\affiliation{Center for Nanostructured Graphene (CNG), Technical University of Denmark, {\O}rsteds Plads, Bldg.~345C, DK-2800 Kongens Lyngby, Denmark}

\author{Kristian S. Thygesen}
\affiliation{Department of Physics, Technical University of Denmark, Fysikvej, Bldg.~307, DK-2800 Kongens Lyngby, Denmark}
\affiliation{Center for Nanostructured Graphene (CNG), Technical University of Denmark, {\O}rsteds Plads, Bldg.~345C, DK-2800 Kongens Lyngby, Denmark}

\author{Martijn Wubs}
\affiliation{Department of Photonics Engineering, Technical University of Denmark, {\O}rsteds Plads, Bldg.~345A, DK-2800 Kongens Lyngby, Denmark}
\affiliation{Center for Nanostructured Graphene (CNG), Technical University of Denmark, {\O}rsteds Plads, Bldg.~345C, DK-2800 Kongens Lyngby, Denmark}

\begin{abstract}
Using a nearest-neighbor tight-binding model we investigate quantum effects of  plasmons on few-nanometer wide graphene nanoribbons, both for zigzag and armchair edge terminations.
With insight from the Dirac description we find an emerging scale-invariant behavior that deviates from the classical model both for zigzag and armchair structures. The onset of the deviation can be related to the position of the lowest parabolic band in the band structure. Dirac theory is only valid in the parameter subspace where the scale invariance holds that relates narrow ribbons with high doping to wide ribbons with low doping. 
We also find that the edge states present in zigzag ribbons give rise to a  blueshift of the plasmon, in contrast to earlier findings for graphene nanodisks and nanotriangles.
\end{abstract}
\maketitle

\section{Introduction}

Since its discovery,\cite{Novoselov2004ElectricFilms} graphene has attracted much attention in the scientific community.
Initially mainly for its remarkable electronic properties as well as its unprecedented mechanical qualities.\cite{Novoselov2012}
However, the plasmonic capabilities of this conveniently tunable material have also received great interest in recent years\cite{GarciadeAbajo2014GrapheneOpportunities,Bonaccorso2010GrapheneOptoelectronics,Grigorenko2012GraphenePlasmonics,Xiao:2016,Goncalves2016AnPlasmonics}
along with other two-dimensional materials.\cite{Mak2016PhotonicsDichalcogenides}
As  ever smaller and more precise devices\cite{Cai2010AtomicallyNanoribbons,Narita2014SynthesisNanoribbons,Narita2015NewChemistry,Ruffieux2016On-surfaceTopology,Wang2016GiantEdges} are produced, it is important to obtain corresponding theoretical understanding of plasmons in graphene nanostructures.
For instance nanodisks and nanotriangles have both been investigated both theoretically\cite{Thongrattanasiri2013OpticalNanostructures,Christensen2014ClassicalStates,Wang2015PlasmonicNanotriangles,Settness:2017} and in experiments,\cite{Wang2016ExperimentalWindow} and more complex structures have also been studied.\cite{Zheng:2017}

The electronic properties of graphene nanostructure can be described on various levels of sophistication. Classically, it is a finite-size conductivity sheet. The simplest atomistic description is a tight-binding (TB) model for the electrons. The Dirac-equation continuum model for finite graphene structures is of intermediate complexity and describes low-energy electrons with linear dispersion being confined on finite graphene structures. Each of these three electronic models has its associated optical response, so that plasmonic excitations may also vary. While the tight-binding model is the most microscopic of them, it is important to know when the simpler Dirac or even the classical description suffices, and for which parameters the three models start to deviate from each other, and how important for optical properties are the different electronic edge terminations.\cite{Zarenia:2011a,Wettstein:2016a}

\begin{figure}[htbp]
\centering
\includegraphics[width=3.5in]{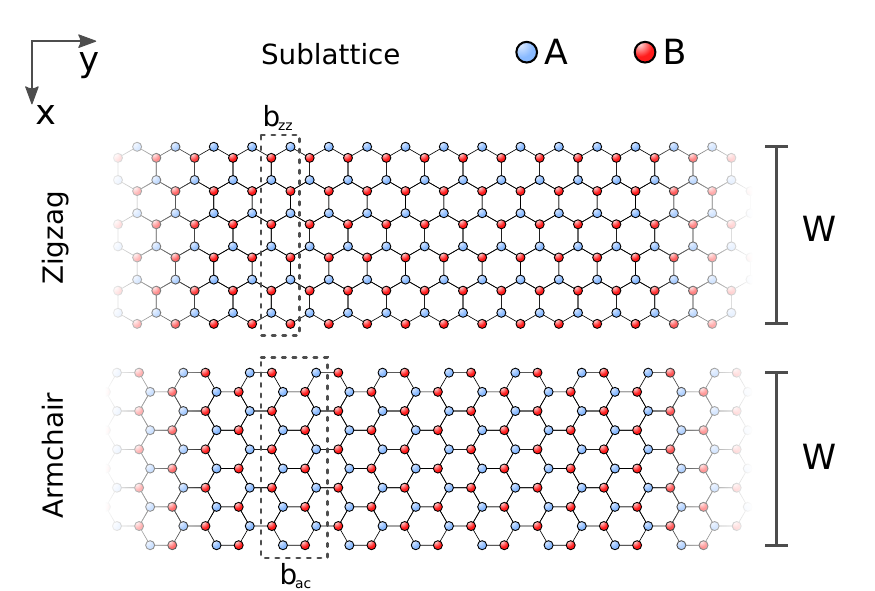}
\caption{The geometries of the zigzag and armchair graphene ribbons. The supercells are marked with the dashed rectangles.}\label{fig:ribbongeometry}
\end{figure}

In this paper we present quantum mechanical calculations of graphene nanoribbons, with geometries as depicted in Fig.~\ref{fig:ribbongeometry}, 
in particular tight-binding calculations in the random-phase approximation (RPA). Important previous work on this topic includes theoretical contributions  both for isolated ribbons and for arrays of them,~\cite{Thongrattanasiri2012QuantumPlasmons,Karimi2017PlasmonsNanoribbons,Shylau2015ElectronNanoribbons,Brey2007ElementaryNanoribbons,Villegas2013PlasmonNanoribbons,Andersen2012PlasmonNanoribbons,Wenger2016OpticalGraphene,Goncalves2016ModelingApproach} as well as experimental studies\cite{Xu2016EffectsNano-imaging,HuImagingNanoribbons,Fei2015EdgeNanoribbons,Brar2014HybridHeterostructures,Ju2011GrapheneMetamaterials,Yan2013DampingNanostructures} with ribbon widths down to 15 nm.\cite{BrarHighlyNanoresonators}
Furthermore, in a complementary analytical analysis we identify a scale invariance in the Dirac-equation model for graphene ribbons,\cite{Brey2006ElectronicEquation} a scale invariance that it shares with the classical model but not the tight-binding model. For the latter we identify the scale invariance as an emergent property. Thereby we obtain an illuminating overview for which parameters the Dirac-equation model can agree with the tight-binding models. Furthermore, we identify a scale invariant onset of quantum mechanical effects.

The article is structured as follows: In Sec.~\ref{sec:methods} we briefly discuss the TB model and its numerical evaluation, and the corresponding optical response function in terms of the electronic states. In Sec.~\ref{Sec:Analytical} we review the Dirac-equation model both for zigzag and armchair graphene ribbons, use the band structures to identify the onset of non-classical effects, and we identify the dimensionless scaling behavior property. In Sec.~\ref{sec:results} we compare our numerical TB calculations with our analytical predictions, and we conclude in Sec.~\ref{Sec:conclusions}. Detailed information can be found in two appendices.

\section{Numerical methods}\label{sec:methods}

\subsection{Tight-binding model}\label{sec:TB}
We describe the graphene ribbon in a nearest-neighbor tight-binding model with the Hamiltonian
\begin{align}
H = \sum_{<i,j>} -t(a^\dagger_ib_j + h.c.),
\end{align}
where the sum is over pairs of neighboring sites.
This model has proven useful for describing the band structure in a wide energy range around the Dirac point as the bands here are determined by interaction between the $p_z$ orbitals of the $sp^2$ hybridized carbon atoms.
A hopping value of $t = 2.8\unit{eV}$ is used between all interacting atoms as it has generally been found to give good results.\cite{CastroNeto2009}

We have used the smallest possible supercell which includes one row of atoms for the zigzag (ZZ) ribbons and two rows for the armchair (AC) ribbons as illustrated in Fig.~\ref{fig:ribbongeometry}.
The band structure and the states are found by direct diagonalization of the Hamiltonian with a $k$-point sampling of at least 5000 points in the Brillouin zone which has been found to give converged results in the subsequent evaluation of the optical response.

\begin{figure}[htbp]
\centering
\includegraphics[width=3.4in]{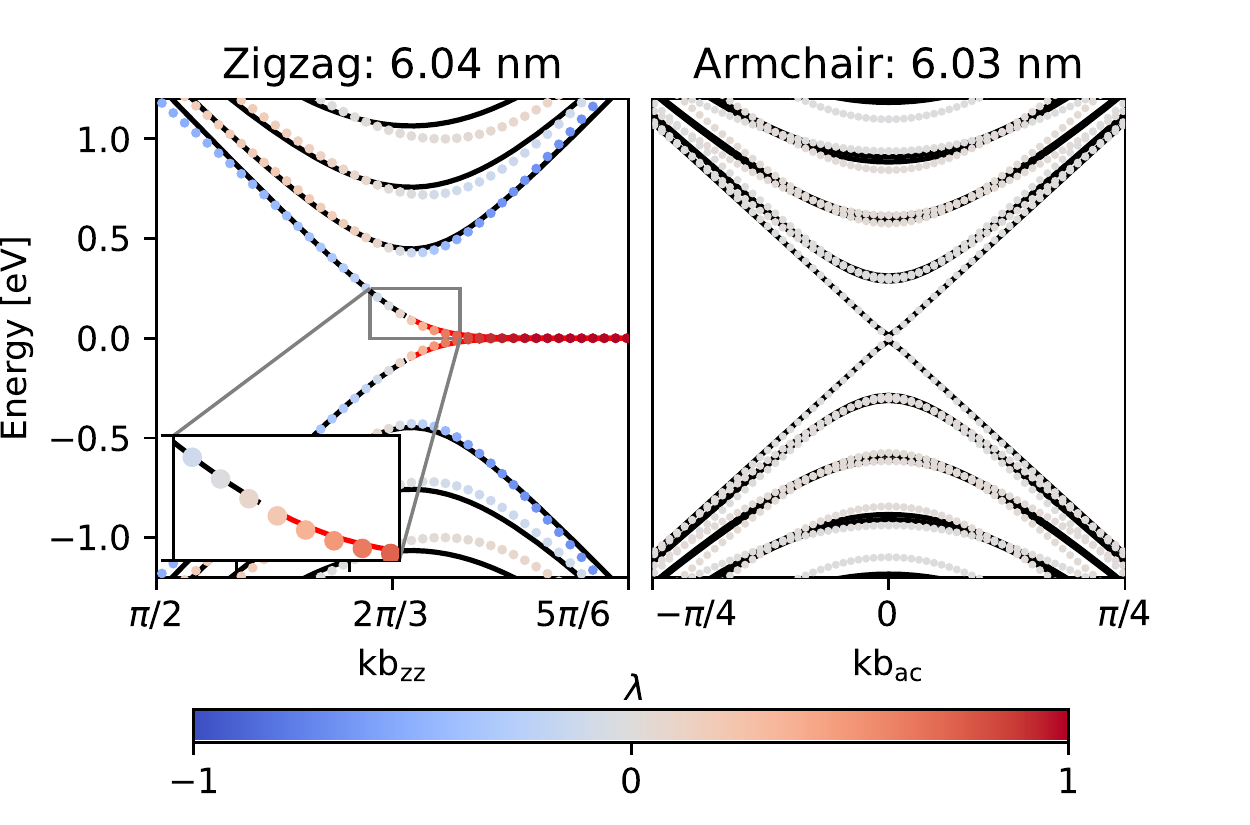}
\caption{(Color online.) Band structures of  6 nm-wide zigzag and armchair ribbons. Full lines are the bands obtained with the Dirac theory, with red color indicating the edge states. The colored dots are the corresponding energies as calculated with the TB model. The color bar indicates how ``edge-like'' the states are, as defined in Eq.~(\ref{Eq:edginess}), with the value $\lambda = 1$ corresponding to a state localized on the edges and -1 to a state localized in the center of the ribbon. The armchair states are uniformly distributed across the ribbon and thus satisfy $|\lambda| \ll 1$ for all states.}\label{fig:bandstructure}
\end{figure}

In Fig.~\ref{fig:bandstructure} we show the bands around the $K$ point for two 6 nm-wide ribbons, one with ZZ and the other with AC edge terminations.
The dots correspond to TB calculations, and the colors indicate the \emph{edginess} (defined in Appendix~\ref{app:edgyness}) of the corresponding states, with bright red corresponding to an edge state.
The figure also shows the continuous bands calculated within  Dirac theory, as discussed in Sec.~\ref{Sec:Analytical}.

\subsection{Response function and quantum plasmons}\label{Sec:Response_function}

The optical response of a quantum mechanical system is described in full by the dielectric function which relates to the non-interacting electron density-density response operator $\chi^0$ in the RPA  as
\cite{Wunsch2006DynamicalDoping,Winther2015QuantumNanostructures}
\begin{align}
\epsilon(\vr,\vr';\omega) = \delta(\vr,\vr') - \Intopnl{\vr''} V(\vr,\vr'')\chi^0(\vr'',\vr';\omega), \label{eq:epsilon_rpa_def}
\end{align}
where $V$ is the Coulomb interaction. Following the method of Ref.~\onlinecite{Thongrattanasiri2012QuantumPlasmons}, the $\chi^0$ is calculated from the TB eigenstates, in the case of only vertical excitations, \ie\ neglecting intraband transitions, as
\begin{align}
\chi_{ij}^0(\omega) = \frac{2e^2}{\hbar} \frac{b}{2\pi}\int\limits^{\mathrm{BZ}}\!\!{\mathrm{d}k} \sum_{nm} f_{nm}\frac{a^{}_{in}a^*_{im}a^*_{jn}a^{}_{jm}}{\epsilon_{nm}+\hbar(\omega + i\eta)}. \label{eq:suscep_def}
\end{align}
Here the $i,j$ run over atomic sites and these indices are thus  discretized equivalents to the $\vr'$ and $\vr''$ coordinates in Eq.~\eqref{eq:epsilon_rpa_def}, while $n,m$ label the eigenmodes at wave vector $k$.
Thus, $a_{in}$ is the weight of the $n^{\rm th}$ wavefunction on the $i^{\rm th}$ site (implicitly at wave vector $k$).
We have used the shorthand notation $\epsilon_{nm} = \epsilon_n - \epsilon_m$ for the energy difference and similarly for the Fermi filling factors $f$.
Damping is included phenomenologically through the parameter $\eta$, which we set to $1.6\unit{meV}$ throughout as in Ref.~\onlinecite{Thongrattanasiri2012QuantumPlasmons}.
The parameter $b$ is the width of the supercell in the periodic direction, see Fig.~\ref{fig:ribbongeometry}.

From the density-density response function we calculate the dielectric matrix $\epsilon(\omega) = I - V\chi^0(\omega)$. This expression involves the Coulomb interaction $V$ in real space, which is a subtle matter to handle,\cite{Thongrattanasiri2012QuantumPlasmons} both due to its long-range behavior and because of the divergence at zero distance, but it can be done (details in App.~\ref{app:coulomb}).
As shown in Ref.~\onlinecite{Andersen2012SpatiallyFirst-principles}, $\epsilon(\omega)$ can be written in a spectral representation of its eigenvalues and left and right eigenvectors as
\begin{align}\label{eq:epsilon}
\epsilon_{ij}(\omega) = \sum_n \epsilon_n(\omega)\phi_{n,i}(\omega)\rho^*_{n,j}(\omega),
\end{align}
where the $i,j$ are again site indices in the tight-binding basis and the $\epsilon_n(\omega)$ the eigenvalues; the right eigenvector $\phi_n$ is the induced field, and the left eigenvector $\rho_n$ is the induced charge of the plasmon. The zeroes of the real parts of $\epsilon_n(\omega)$ define the plasmonic modes.   The plasmonic modes thus found agree well with peaks in the energy-loss function $-\Im \epsilon^{-1}(\omega)$ as  measured in electron energy loss spectroscopy experiments, provided the frequency dispersion of the imaginary part of $\epsilon_n(\omega)$ is small. The above method to calculate quantum plasmons based on a tight-binding formalism will be applied to graphene ribbons in Sec.~\ref{Sec:Response_function}.

\section{Analytical model}\label{Sec:Analytical}

\subsection{Dirac theory for graphene ribbons}
Dirac theory is an approximate theory obtained 
by linearizing the TB model in the $K$ ($K'$) valleys where infinite graphene exhibits its Dirac cones.
This  allows one to get analytical insight into the band structure also of finite graphene structures. 
For graphene ribbons, this was first done in the seminal paper by~Brey and Fertig\cite{Brey2006ElectronicEquation} and the method is also outlined in Castro Neto et al.~\cite{CastroNeto2009}
Here we first briefly review the Dirac theory, before presenting our new analytical insights and their comparison with full TB calculations. 

In its essence, in the low-momentum limit the tight-binding Hamiltonian is approximated as 
\begin{align}
H &= \hbar v_F (\tau_0 \otimes \sigma_x k_x + \tau_z \otimes \sigma_y k_y) \label{eq:hamilton}\\
&=\hbar v_F\begin{pmatrix}
0 & k_x - ik_y & 0 & 0 \\
k_x + ik_y & 0 & 0 & 0\\
0 & 0 & 0 & -k_x - ik_y\\
0 & 0 & -k_x + ik_y & 0
\end{pmatrix}, \notag
\end{align}
where $\tau_i$ and $\sigma_i$ are the Pauli spin-matrices belonging to the valley space and sub-lattice space, respectively.
For the eigenstates of the system we adopt the notation of Ref.~\onlinecite{Brey2006ElectronicEquation}: $[\phi^A,\phi^B,-\phi^{A'},-\phi^{B'}]^T$.
The Hamiltonian in Eq.~\eqref{eq:hamilton} is block diagonal, so we focus only on the upper left corner corresponding to the $K$~valley.
By applying $H$ twice to a state $[\phi^A,\,\phi^B]^T$ we find the relations
\begin{subequations}\label{eq:diraccoupledequations}
\begin{align}
\pa{k_x^2+k_y^2} \phi^A &= \tilde\epsilon_k^2 \phi^A \\
\pa{k_x^2+k_y^2} \phi^B &= \tilde\epsilon_k^2 \phi^B,
\end{align}
\end{subequations}
with $\teps = \epsilon/\hbar v_F$.

Now we specify the nanostructure to be a graphene ribbon in the $xy$-plane that is infinite in the $y$-direction and has width W. When Fourier transforming Eq.~(\ref{eq:diraccoupledequations}) in only the $x$-direction, \ie\ by replacing $k_x$ with $-i\partial_x$, a differential equation is obtained with the general solutions
\begin{align}\label{Eq:general_form_solutions}
\phi^X(x) = Ae^{\beta x} + Be^{-\beta x},
\end{align}
with $\beta = \sqrt{\smash[b]{k_y^2 - \teps^2}}$, and consequently $\epsilon = s\hbar v_F \sqrt{\smash[b]{k_y^2 - \beta^2}}$ where $s=\pm1$.
The eigenmodes of the Hamiltonian can be found analytically for both possible ribbon geometries of  Fig.~\ref{fig:ribbongeometry} by imposing  proper corresponding boundary conditions for their wavefunctions. These boundary conditions are different for zigzag and for armchair edge terminations.

\subsubsection{Zigzag edge termination}\label{sec:Zigzag}
In a ZZ ribbon the atomic structure terminates on an $A$ lattice site on one edge and on a $B$ site on the opposite edge, see Fig.~\ref{fig:ribbongeometry}. The proper boundary conditions are that $\phi^A(x=0) = \phi^B(x=W) = 0$. 
By applying these boundary conditions to the general solution~(\ref{Eq:general_form_solutions}), the dispersion relation for the allowed states is found.
In a slightly different notation than in Ref.~\onlinecite{Brey2006ElectronicEquation}, it reads
\begin{align}
k_y = \frac{\beta}{\tanh(\beta W)}.\label{eq:tran_eq}
\end{align}
For fixed $k_{y}$, Eq.~\eqref{eq:tran_eq} has infinitely many solutions for imaginary $\beta=ik_n$ corresponding to the bulk modes, and at most one solution for $\beta = \kappa\in\mathbb{R}$ corresponding to an edge mode that falls off exponentially fast away from the edge.
It follows from the limit $\lim_{\kappa\to0} \kappa/\tanh(\kappa W) = 1/W$ that the edge states only exist for $k_y\geq 1/W$. This momentum cut-off has an associated energy cutoff $\varepsilon_\mathrm{cut} = \hbar v_F/W$.
  
The two types of solutions (bulk and edge modes) are shown in Fig.~\ref{fig:bandstructure} as full lines in black and red, respectively.
It is clear from the figure that the TB and Dirac methods to calculate the band structure give very similar energies in the vicinity of the $K$ point and that the analytically found edge states match almost perfectly with the ``edgy'' ($\lambda \approx 1$) states in TB.
From the analytical model we just determined the exact energy range where the edge states are found.
Given the great agreement between the two approaches, in the following, where we want to distinguish between bulk and edge states, we use the energy cut-off $\varepsilon_\mathrm{cut}$ from Dirac theory to classify the TB states as either bulk- or edge-like.

\subsubsection{Armchair edge termination}

As the termination of an armchair ribbon has a mix of $A$ and $B$ lattice sites, as depicted in Fig.~\ref{fig:ribbongeometry}, we demand that the sublattice wavefunction vanishes on both edges.
This results in a mixing of $K$ and $K'$ states through the equations\cite{CastroNeto2009}
\begin{align}
\begin{split}
0 &= \phi^{A/B}(x=0) + \phi^{A'/B'}(x=0), \\
0 &= e^{iKW}\phi^{A/B}(x=W) + e^{-iKW}\phi^{A'/B'}(x=W),
\end{split}
\end{align}
where $K = 4\pi/3\sqrt{3}a_0$ and $-K$ are the positions of the $K$-valleys in momentum space and $a_0$ is the interatomic distance in the graphene lattice.
These boundary conditions together with the general form of the solution~(\ref{Eq:general_form_solutions}) yield plane-wave states of the form $e^{ik_n x}$ with $k_n$ given by\cite{Brey2006ElectronicEquation}
\begin{align}
k_n = \frac{n\pi}{W} - \frac{4\pi}{3\sqrt{3}a_0} = \frac{2\pi[3n-2(N+1)]}{3\sqrt{3}a_0(N+1)},
\end{align}
with $n\in\mathbb{Z}$, and the corresponding eigenenergies  $\teps_{n} = s\sqrt{\smash[b]{k_y^2+k_n^2}}$.
In the second equality we have expressed the width of the ribbon as $W = (N+1)a_0\sqrt{3}/2$, where $N$ is the number of atomic rows.
From this form it follows that every third ribbon, where $3n-2(N+1) = 0$ can be fulfilled, will be semi-metallic while the rest will have a band gap.
This way of defining the width gives the correct order of semi-metallic and semiconducting ribbons when compared with tight-binding calculations.
For example, the narrowest ribbon has $N=2$ and is just a one-dimensional chain with corresponding cosine band structure and no band gap. There should therefore exist a solution of the above condition for a semi-metal, $3n-2(2+1) = 0$, which is the case for $n=2$.

\subsection{Dimensionless scaling in Dirac theory}\label{Sec:Dirac_scaling}
An important property of the Dirac theory is a scale invariance of the ribbons:
if all equations are rewritten in dimensionless units where the energies are scaled in units of the Fermi energy $\varepsilon_F$,  momenta in units of the Fermi momentum $k_F$,  and the distances with the ribbon width $W$, then one finds that the only system-dependent parameter is  the dimensionless parameter $\Lambda \equiv k_F W$.
This insight is very useful, since it allows us to identify effects that should exist across all widths of ribbons, provided that their respective Fermi levels are scaled accordingly and provided of course that the Dirac model is valid. 

\begin{figure}
\centering
\includegraphics[width=3.4in]{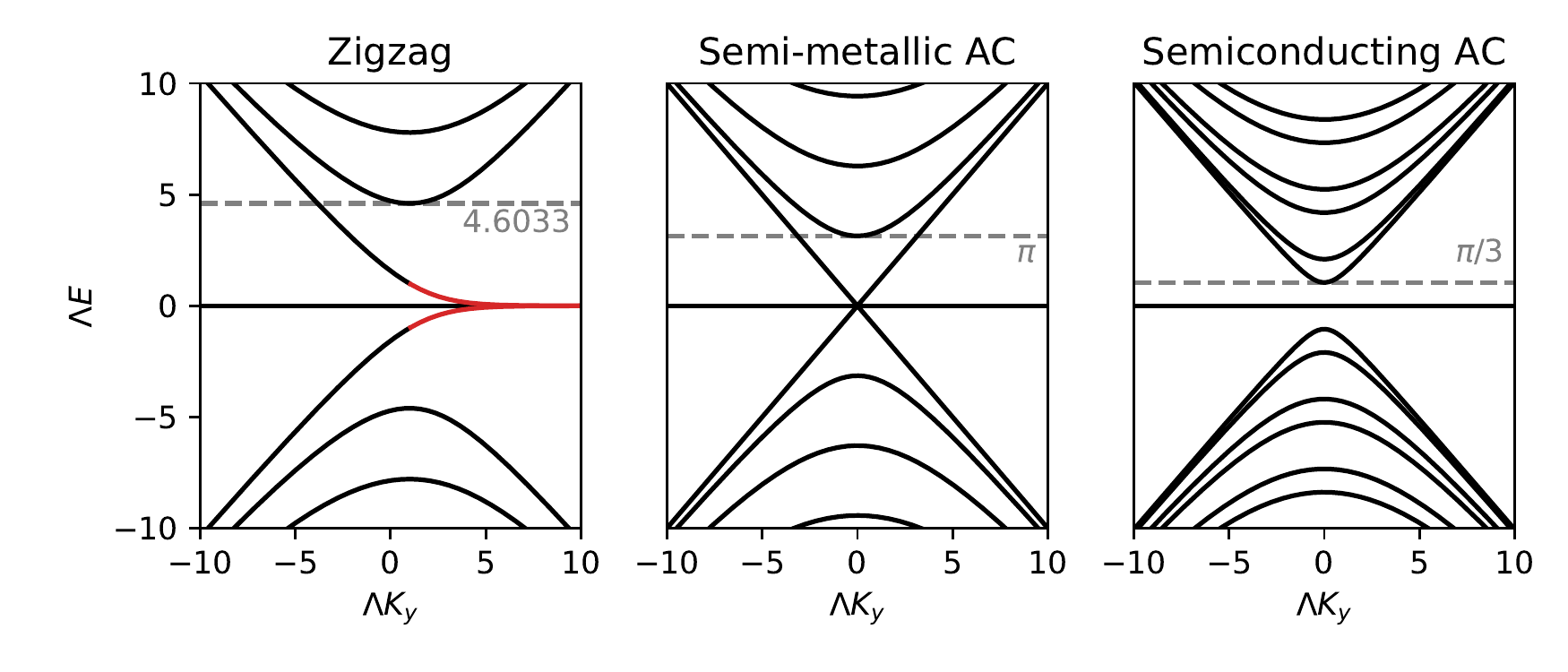}
\caption{(color online) The band structures of zigzag, semi-metallic, and semiconducting ribbons in dimensionless units. We expect non-classical behavior when the Fermi energy is close to or below the bottom parabolic band indicated with the gray dashed lines.}\label{fig:dimensionless_bands}
\end{figure}

In dimensionless form, the governing equations for the ZZ ribbons thus become
\begin{subequations}
\begin{align}
K_y &= \frac{K_n}{\tan\pa{K_n\Lambda}}, & K_y &= \frac{K}{\tanh\pa{K\Lambda}} \\
E_n &= \sqrt{K_y^2+K_n^2}, & E_e &= \sqrt{K_y^2 - K^2} \\
\psi(\tilde x) &= \mathrlap{C_b e^{iK_y\Lambda\tilde y}
\mathrlap{\begin{pmatrix}
is \sin\pa{\tilde x\Lambda K_n} \\
\sin\pa{\pb{1-\tilde x}\Lambda K_n}
\end{pmatrix}}} \\
\phi(\tilde x) &= C_e e^{iK_y\Lambda\tilde{y}}
\mathrlap{\begin{pmatrix}
is \sinh\pa{\tilde{x}\Lambda K} \\
\sinh\pa{\pb{1-\tilde x}\Lambda K}
\end{pmatrix},}
\end{align} \label{eq:dimensionless_form}
\end{subequations}
with dimensionless momentum $K_y \equiv k_y/k_F$ and corresponding dimensionless momentum and energy of the bulk modes $\psi(\tilde x)$ are denoted by $K_n$ and $E_n \equiv \epsilon_n/\varepsilon_F$, and those for the edge modes $\phi(\tilde x)$ are called $K$ and $E_e$, and  $\tilde x = x/W$ is the dimensionless lateral position in the ribbon.
For the AC ribbons we  find that
\begin{align}
K_n = \frac{\pi[3n - 2\pa{N+1}]}{3\Lambda}.
\end{align}
Plots of the dimensionless band structures for the three different cases, zigzag, and semi-metallic and semiconducting armchair ribbons are shown in Fig.~\ref{fig:dimensionless_bands}.
We  emphasize the large differences between the band structures in the low-energy regime; especially the different placement of the bottom of the lowest parabolic band, to which we will return in the following.

One important prediction is that the scaling behavior of the electronic states will carry over to the plasmonic energies as well, as can be seen by inserting the Dirac states in their dimensionless form into the density-density response function, Eq.~\eqref{eq:suscep_def}, here shown with the bulk states:
\begin{align}\label{eq:X0scaleinvariantbulkonly}
\chi^0 \propto \frac{1}{\Lambda}\Intop{0}{\infty}{K_y}\sum_{nm}f_{nm}\frac{\abs{C^{}_bC'_b}S_{nm}(\tilde x, \tilde x', \Lambda)S_{mn}(\tilde x, \tilde x', \Lambda)}{E_n - E_m - (\nu -i\gamma)}
\end{align}
with
\begin{multline}
S_{nm}(x,x',\Lambda) = ss'\sin\pa{\tilde x\Lambda K_n}\sin\pa{\tilde x' \Lambda K_m} \\
-\sin\pb{(1-\tilde x)\Lambda K_n}\sin\pb{(1-\tilde x')\Lambda K_m},
\end{multline}
where $\nu + i\gamma = \hbar(\omega + i\eta)/\varepsilon_F$.
As the band structure in the Dirac model is approximated with an infinite cone the (dimensionless) response will only depend on $\Lambda$, rather than on $\varepsilon_{F}$ and $W$ separately. This scale invariance holds both when leaving out the edge states, as in Eq.~(\ref{eq:X0scaleinvariantbulkonly}), and when including them.

Given the great agreement in the band structures of the numerical tight-binding and the analytical Dirac models, we expect that also for TB calculations there will be parameters for which the scaled plasmonic energies will be scale-invariant for constant values of the parameter $\Lambda$, even though the scale invariance does not strictly hold in the TB model.

\subsection{Emergence of non-classicality}\label{sec:emergence_of_nonclassicality_predicted}

The quantum mechanical Dirac model for ribbons has a classical limit, and vice versa away from the classical  limit we will identify the onset of non-classical behavior.
When there are many bands crossing the Fermi energy one would expect the system to behave classically.
On the other hand, for combinations of widths and Fermi energies where the Fermi surface is only crossed by a few states we are starting to probe the quantumness of the system and expect deviations from the classical regime.
As our heuristic measure, we take the bottom energy of the lowest parabolic band as the separation between the quantum and classical regimes.
Interestingly, from Fig.~\ref{fig:dimensionless_bands} this value differs for ZZ and AC ribbons, and it differs also for the semi-metallic and the semiconducting AC ribbons. 
These different critical values $\Lambda_{c}$ at which we predict the classical-to-quantum behavior to occur can be determined analytically (and further below we will test them against numerical TB calculations).
\paragraph{For zigzag:}
By setting to zero the derivative of the energy with respect to the dimensionless momentum $K_y$, it is found that the sought bottom of the band occurs at $K_y=1/\Lambda$, corresponding to $k_y=1/W$.
By inserting this into the scaled expression for the band energies, we find that
\begin{align}
E_n = \sqrt{\Lambda^{-2}+K_n^2(\Lambda^{-1})} = \frac{\sqrt{1+\xi_n^2}}{\Lambda},
\end{align}
where $\Lambda K_n= \xi_n = \tan(\xi_n)$.
Looking for the solution where the Fermi energy crosses the bottom parabolic band, \ie\ $\epsilon_n/\varepsilon_F = E_n = 1$, it is found that the critical value is
\begin{align}
\Lambda_\mathrm{c}^{\rm zz} &= \sqrt{1 + \xi_1^2} \approx 4.6033. \label{eq:Lambda_critical_zz}
\end{align}
This $\Lambda_\mathrm{c}^{\rm zz}$ is a dimensionless number, and with this single number we predict with Dirac theory the emergence of quantum effects both in narrow ribbons at high Fermi levels and in  wide ribbons with low doping.
As we will see below, this is indeed the value around which the dipole plasmon energies start to deviate from the classical results for zigzag ribbons.

\paragraph{For armchair:} For ribbons with armchair edge terminations, in the limit of many atoms, the band bottoms occur at $E_n = \pm n\pi/\Lambda$ for the semi-metallic ribbons and at $E_n \in \{\pm(3n+1)\pi/3\Lambda, \pm(3n+2)\pi/3\Lambda\}$ for the semi-conducting ribbons, with $n\in\mathbb{N}$.
\Ie\ 
\begin{align}
\Lambda_\mathrm{c}^{\rm ac} =
\begin{cases}
\pi & \text{for semi-metallic AC ribbons}\\
\frac{\pi}{3} & \text{for semiconducting AC ribbons}
\end{cases}, \label{eq:Lambda_critical_ac}
\end{align}

Unlike for the ZZ ribbons, the band structures for AC ribbons are symmetric around the Dirac points and in that sense they are thus more like the bulk graphene bands. Combined with the lower value of $\Lambda_\mathrm{c}^{\rm ac}$,  we expect classical behavior down to smaller values of $\Lambda$ for armchair ribbons.

\section{Classical plasmons}\label{sec:classical_plasmons}
It is naturally also possible to calculate the plasmons classically.
For the ribbon geometry this has already been done  in different ways.\cite{Thongrattanasiri2012QuantumPlasmons,Christensen2015FromDimensions,Velizhanin2015GeometricArrays,Goncalves2016AnPlasmonics}
When combined with the continuity equation, the coupling between the potential $\phi(\vr)$ and the induced charge density $\rho(\vr)$ can be written as an integro-differential eigensystem of equations as
\begin{subequations}\label{eq:IDE_for_classical_plasmons}
\begin{align}
\zeta_n\phi_n(\vr) &= \frac{-1}{2\pi}\Intopnl{{}^2\vr}\frac{\nabla'\cdot\pb{f(\vr')\nabla'\phi_n(\vr')}}{\abs{\vr-\vr'}}, \\
\zeta_n &=\frac{2i\epsilon_0\omega_nW}{\sigma(\omega_n)}, \label{eq:Classical_IDE_eigenvalue}
\end{align}
\end{subequations}
where all coordinates and differential operators work in the 2D plane of the graphene.
The graphene is treated as being embedded in an $\epsilon=1$ material.
It has here been assumed that the conductivity is uniform inside the ribbon of width $W$, and vanishes outside: 
\begin{align*}
\sigma(\vr,\omega) &= \sigma(\omega)f(\vr), \qquad\text{with} \\
f(\vr) &= \begin{cases}
1 \qquad \text{for $\vr$ inside the ribbon},\\
0 \qquad \text{for $\vr$ outside the ribbon}.
\end{cases}
\end{align*}

High-precision fits of the values of the eigenvalues $\zeta_n$ in Eq.~(\ref{eq:IDE_for_classical_plasmons}) are given by Christensen (Ref.~\onlinecite{Christensen2015FromDimensions}) for the first seven modes.
We have used these values in our classical calculations together with the low-temperature, local conductivity $\sigma(\omega)$ for bulk graphene. This conductivity can be derived, among other ways, from the Dirac model in the limit of infinitely wide ribbons or from the general expression of the bulk polarizability of graphene as found by Hwang and Das Sarma,\cite{Hwang2007DielectricGraphene} and by Wunsch \textit{et al.} \cite{Wunsch2006DynamicalDoping}
Here we just present the resulting expressions for the intraband and the interband contributions that together make up $\sigma(\omega)$:
\begin{subequations}\label{eq:sigma_intra_inter}
\begin{align}
\sigma_\mathrm{intra}(\omega) &= \frac{ie^2\varepsilon_F}{\pi\hbar^2(\omega+i\eta)}, \\
\sigma_\mathrm{inter}(\omega) &= \frac{e^2}{4\hbar}\pb{\frac{i}{\pi}\ln\abs{\frac{2\varepsilon_F - \hbar\omega}{2\varepsilon_F+\hbar\omega}} + \Theta(\hbar\omega - 2\varepsilon_F)},
\end{align}
\end{subequations}
where $\Theta$ is the Heaviside step function.
By combining Eqs.~(\ref{eq:IDE_for_classical_plasmons}) and (\ref{eq:sigma_intra_inter})  we can find the plasma energies as a function of the ribbon width.

For our purposes it is essential to realize that Eq.~\eqref{eq:Classical_IDE_eigenvalue} can be rewritten in dimensionless variables as $\sigma(\nu_n)\zeta_n = 2i\epsilon_0\hbar v_F\nu_n\Lambda$, with the dimensionless plasmon energy $\nu_n = \hbar\omega_{n}/\varepsilon_F$ and again $\Lambda = k_F W$.
This insight turns out to be quite practical, because it is sufficient to calculate the connection between $\nu_n$ and $\Lambda$ only once to obtain the plasmon energies for all combinations of widths and Fermi momenta that satisfy $\Lambda = k_FW$. Moreover, in Sec.~\ref{Sec:Dirac_scaling} we saw that the Dirac model has the same scale invariance. So we find that the scaling property  holds both inside and outside the classical regime, as long as Dirac theory is accurate. We will test the latter by comparing Dirac and classical theories with tight-binding calculations in the next section.

\section{Numerical and analytical results compared}\label{sec:results}

We present two comparisons: quantum versus classical plasmons in  Sec.~\ref{sec:quantumclassical}, and properties of atomistic (TB) quantum plasmons versus those of continuum (Dirac) quantum plasmons in Sec.~\ref{sec:emergentscaleinvariancecomparison}.  

\begin{figure*}
\centering
\subfloat[c][Zigzag]{\includegraphics[width=3.4in]{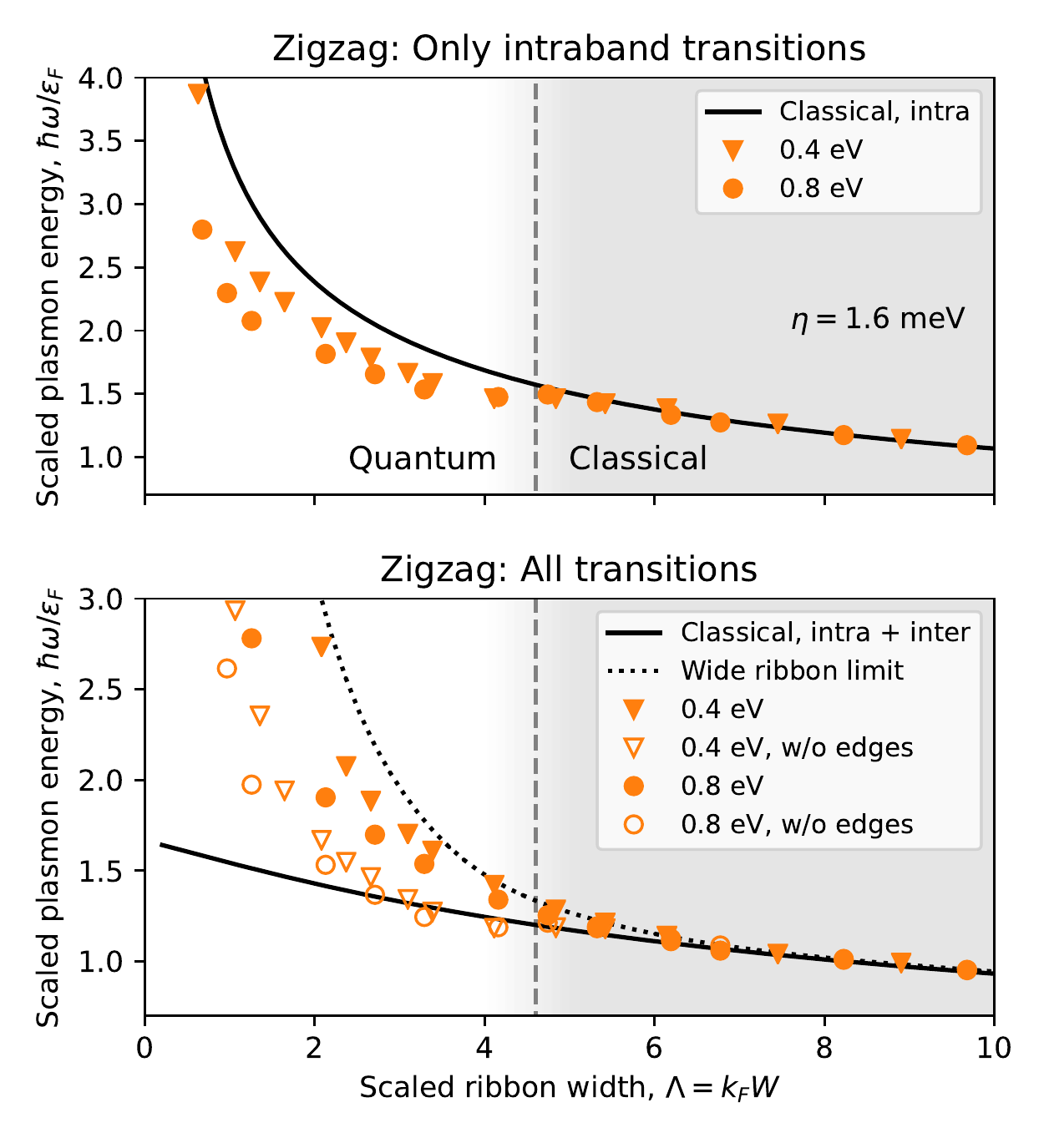}\label{fig:widths_zz}}
\subfloat[c][Armchair]{\includegraphics[width=3.4in]{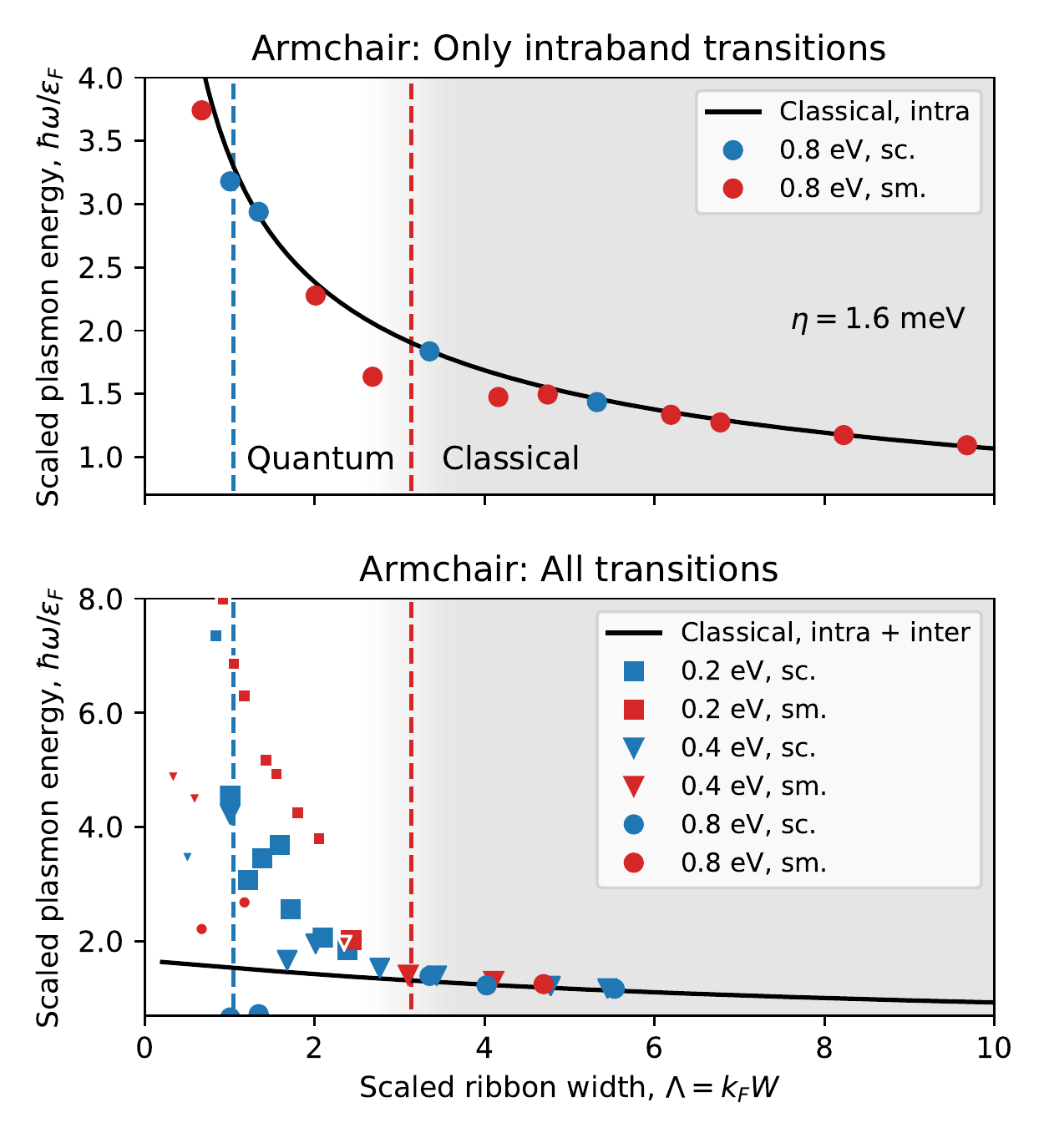}\label{fig:widths_ac}}
\caption{(Color online) 
Scaled plasmon energy as a function scaled ribbon width $\Lambda$. The value $\Lambda = 5$ corresponds for example to a width of $16.4\unit{nm}$ for $\varepsilon_F = 0.2 \unit{eV}$. 
\emph{(a):} $\Lambda$ is varied by changing $W$ while keeping the Fermi energy fixed at either 0.4~eV (triangles) or 0.8~eV (dots). The vertical dashed line corresponds to $\Lambda_\mathrm{c}^\mathrm{zz}\simeq 4.6$. 
The open symbols in the lower panel correspond to calculations of plasmon energies where edge states were removed from the calculation.
\emph{(b), upper panel:} The intraband plasmons of AC ribbons, both the classical prediction and the quantum plasmon predictions for semiconducting and for semimetal ribbons. Blue and red dashed vertical lines correspond to $\Lambda_\mathrm{c}^{ac} = \pi/3$ and $\pi$, respectively. \emph{(b), lower panel:} As in the upper panel, but now including all transitions. The small symbols are used for the peaks in the loss spectrum that are not associated with an actual plasmon defined as $\mathrm{Re}(\epsilon_n) = 0$.
}\label{fig:contribs_and_widths}
\end{figure*}

\subsection{Quantum versus classical plasmons}\label{sec:quantumclassical}

Here we systematically investigate the range of validity of the classical description for graphene ribbons, by comparing with TB quantum calculations. In particular, we will test the heuristic value of the various $\Lambda_\mathrm{c}$ that we identified in Sec.~\ref{sec:emergence_of_nonclassicality_predicted} for characterizing the emergence of non-classical behavior in a scale invariant way.
In Sec.~\ref{Sec:Response_function} we outlined how one can identify quantum plasmons of nanostructures within a tight-binding formalism, and here we apply this approach to graphene ribbons. The calculation of the corresponding classical plasmons was described in  Sec.~\ref{sec:classical_plasmons}.  

Guided by the scaling properties of the Dirac and classical models,  in Fig.~\ref{fig:contribs_and_widths} we present the plasmonic energies as a function of the dimensionless variable $\Lambda$.

The figure shows a comparison of the scaled plasmonic energy as calculated with the TB model of Sec.~\ref{sec:TB} and in the classical model for both ZZ and AC ribbons, and when considering only the intraband contribution (upper panels of Fig.~\ref{fig:contribs_and_widths})  or all transitions (lower panels).
By ``intraband'' we mean that we only include eigenstates with energies above the cut-off energy $\varepsilon_\mathrm{cut} = \hbar v_F/W$ for the edge states for zigzag ribbons and above zero energy for armchair ribbons, which  corresponds to only considering intraband transitions in a classical, wide-ribbon limit.
For ribbons of finite widths, the transitions are intraband transitions in the sense that the bands in the upper cone are size-constriction-foldings of the infinite graphene conduction band, although the actual transitions do occur between bands of the ribbon.

We see that for large values of $\Lambda$, the classical and all TB calculations agree across all four panels.
There is no visible effect of neither edge terminations nor other quantum effects there.
Furthermore, the TB calculations for different Fermi levels agree very well as predicted from the scaling of the Dirac model. 
For smaller values of $\Lambda$  the plasmon energies as calculated by the TB model start to depart from the classical values.

For zigzag ribbons, Fig.~\ref{fig:contribs_and_widths}(a) constitutes a confirmation of our prediction in Eq.~(\ref{eq:Lambda_critical_zz}) that this onset of quantum behavior occurs at  $\Lambda_\mathrm{c}\simeq 4.6$, the point at which the lowest of the parabolic bands of the zigzag ribbons crosses the Fermi level. This same onset is seen both in the ``Drude-like'' case (Fig.~\ref{fig:contribs_and_widths}(a), upper panel) and with all transition included (Fig.~\ref{fig:contribs_and_widths}(a), lower panel).

Another important feature of Fig.~\ref{fig:contribs_and_widths}(a) is that  the tight-binding plasmon energies for $\varepsilon_{F} = 0.4\unit{eV}$ and $0.8\unit{eV}$ are indeed quite close to each other in the chosen dimensionless units, and closer to each other than to the classical plasmon curves. Dirac theory predicts that the two quantum plasmon calculations would coincide exactly, and the tight-binding calculations confirm that the scale invariance of Dirac theory indeed holds approximately. A better overview and insight when scale invariance holds in TB calculations will be presented in Sec.~\ref{sec:emergentscaleinvariancecomparison} below.
The dotted line in the lower panel shows the interpolated data from calculations of a $9\unit{nm}$ wide ribbon at varying Fermi energy and provides the best guess, given the calculations that have been done, of the behavior of arbitrarily wide ribbons where the plasmon energies have converged with respect to the number of bands.
This will be explored further in the following section.
Comparing the results for the $\varepsilon_F = 0.4\unit{eV}$ and $\varepsilon_F = 0.8\unit{eV}$ ribbons we see that lowering the Fermi energy, which for constant $\Lambda$ corresponds to widening the ribbons, moves the points closer to the dotted line, as expected.

\begin{figure}
\centering
\includegraphics[width=3.5in]{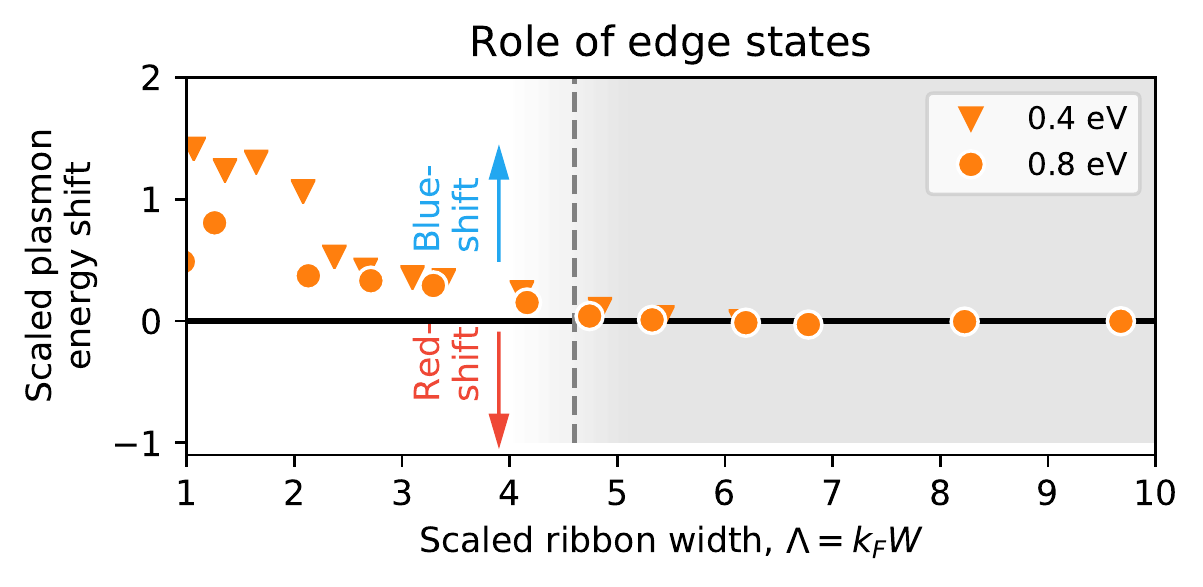}
\caption{(Color online) Scaled plasmon energies in the presence of edge states minus scaled plasmon energies when neglecting the edge states, as a function of the scaled ribbon width $\Lambda$, for two fixed values of the Fermi energy.  }\label{fig:role_of_edges}
\end{figure}

By excluding the zigzag edge states in the evaluation of $\chi^0$ (open symbols in the lower panel of Fig.~\ref{fig:contribs_and_widths}\subref{fig:widths_zz}), we find a significant plasmon redshift in the quantum regime. In other words, edge states of zigzag nanoribbons contribute with a significant blueshift of the plasmon energies in the quantum regime, while they have hardly any impact on the energy in the classical regime above $\Lambda_\mathrm{c}^\mathrm{zz}$.
This effect of edge states becomes even more evident by directly plotting the energy shift as in Fig.~\ref{fig:role_of_edges}. Clearly, for zigzag ribbons the edge states do not affect the plasmon energies for $\Lambda>\Lambda^\mathrm{zz}_\mathrm{c}$ and give rise to a blueshift for $\Lambda<\Lambda^\mathrm{zz}_\mathrm{c}$.
The found blueshift is in stark contrast to the results for graphene disks\cite{Christensen2014ClassicalStates} and triangles\cite{Wang2015PlasmonicNanotriangles} in which the zigzag edge states are found to give rise to a net redshift of the plasmon energies.
Back to our ribbons, for $\Lambda<1$ the Fermi level crosses the edge state and the evaluation of the edge-state contribution in the manner described above becomes meaningless.

Having discussed quantum-classical transitions for zigzag ribbons, we now return to Fig.~\ref{fig:contribs_and_widths} and study armchair ribbons in part~(b).
The picture is slightly different for armchair ribbons as they exist as either semiconducting (sc.) or semi-metallic (sm.).
When including only intraband transitions, the scaled plasmon energies  follow the classical behavior rather closely across the entire range, except for a single outlier.
As discussed above, because of the symmetry around the $K$ point of the armchair band structure, we do not expect the same kind of quantum-classical transition as for zigzag ribbons.
In the lower panel of Fig.~\ref{fig:contribs_and_widths}\subref{fig:widths_ac} we have split the ribbons into the two types.
The vertical, dashed lines indicate the position of the band bottom in the appropriate color.
As expected, the deviation from classical results starts at lower $\Lambda$ than previously for the zigzag ribbons.
The small symbols in the lower panel of Fig.~\ref{fig:contribs_and_widths}(b) denote peaks in the loss spectrum that are not associated with  real plasmons as there is no simultaneous crossing of the real part of the dielectric eigenvalues with zero.
For the semiconducting ribbons the plasmon cease to exist when the Fermi energy crosses the lowest parabolic band at $\Lambda=\pi/3$.
For the semi-metallic ribbons the plasmons  cease to exist earlier, namely already below $\Lambda = \pi$.
There seem to be an exception with the red square just above $\Lambda=2$ (which lies beneath a small, red triangle), but as the TB calculations are done for room temperature $k_bT\approx 25\unit{meV}$, there will still be a finite population of electrons in the bottom parabolic band for this point.
For the smallest values of $\Lambda$ for which plasmons still exist, the positions of the main dipole plasmon peaks become increasingly hard to locate, resulting in an increased scatter of the data points, as also reported elsewhere.~\cite{Christensen2014ClassicalStates,Thongrattanasiri2012QuantumPlasmons}

\subsection{Emergent scale invariance for plasmons}\label{sec:emergentscaleinvariancecomparison}

\begin{figure*}[htbp]
\centering
\includegraphics[width=6.5in]{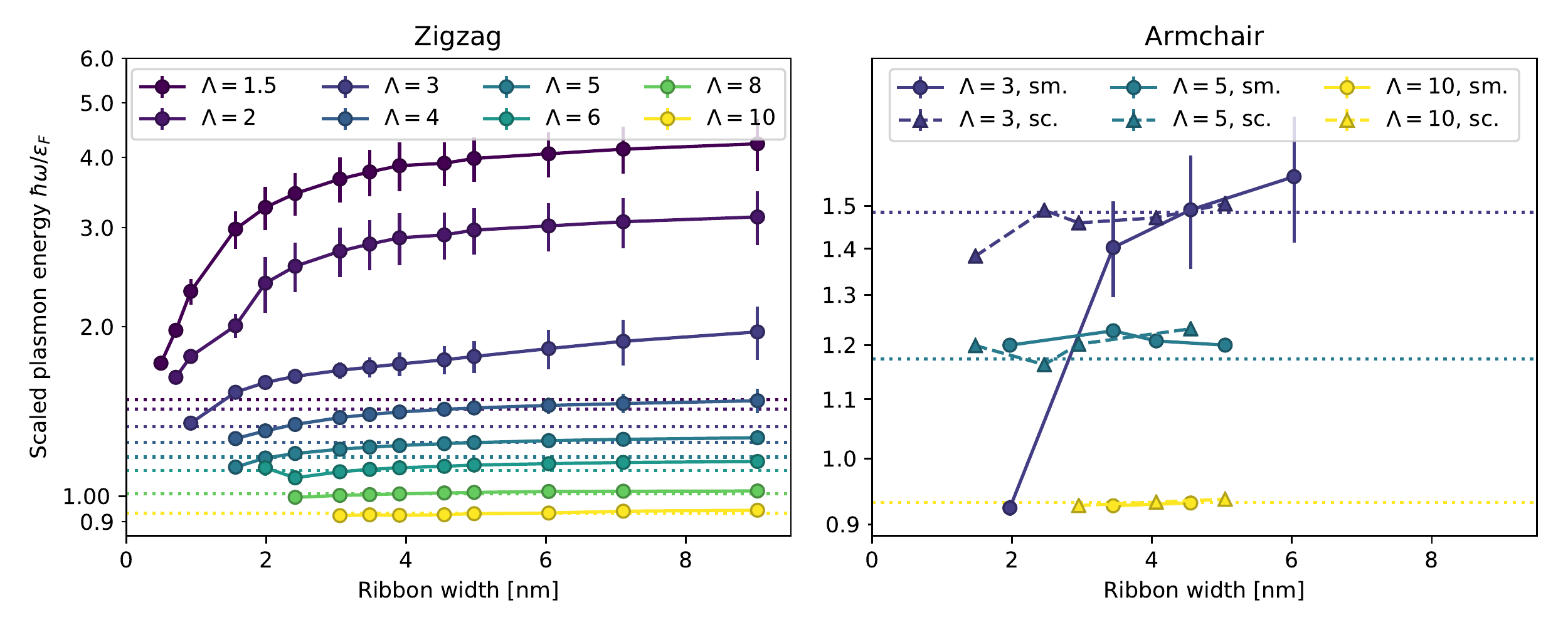}
\caption{(Color online) The plasmon energy scaled with the Fermi energy for constant $\Lambda$ at varying ribbon widths, for both ZZ (panel a) and AC ribbons (b). Dotted horizontal lines in equal colors are the corresponding classical plasmon energies.  In the AC panel the data is split into semi-metallic (sm.) and semiconducting (sc.) ribbons. This distinction is only important for the low $\Lambda$ structures, as seen. The bars show the width of the plasmon peak in the loss spectrum. All displayed data points have $\varepsilon_F < 2.0\unit{eV}$.}\label{fig:constant_lambda}
\end{figure*}
In general, the tight-binding model for graphene ribbons does not have the same scale invariance that we found both for Dirac theory and for classical plasmonics, as the TB band structure does not consist of an infinite Dirac cone.
This follows from the fact that, due to the infinite cone shape, the band structures for two different ribbon widths in the Dirac description are related be a simple scaling transformation while this is not the case for the more complex TB band structure.
But since the low-energy bands calculated with TB and with Dirac theory agree so well, at least for the parameters of Fig.~\ref{fig:bandstructure}, the scale invariance will be an emergent property of the TB model, valid only in part of the parameter space spanned by $\{\varepsilon_F, W\}$.
Only in that subspace can classical and/or Dirac theory be expected to agree with TB calculations.

As a test of the proposed scale invariance we conduct a range of calculations where $\Lambda$ is held constant while the widths of the ribbons are varied, so  doubling the size of the ribbon goes hand in hand with halving  the Fermi energy.
As previously stated, we expect the scaled plasmon energy to tend towards a constant when the ribbons get wider and the Dirac model becomes a better description. It is less clear how fast the limit will be reached. 
When the Fermi energy is above $2.0\unit{eV}$ we are well out of the linear regime of the bands and do not expect the Dirac scaling to work anymore.
For the armchair ribbons we distinguish between semiconducting and semi-metallic ones, as this should have an impact for small values of $\Lambda$ where the Fermi energy is close to the difference in the band structures.

As one of our main results we present in Fig.~\ref{fig:constant_lambda}  how the TB plasmon energies converge as ribbon widths are increased. For $\Lambda \gg 1$ the plasmon energies quickly converge for larger widths to a value that differs little from the classical plasmon energy.
But it is important to notice that the wide-ribbon limits in this figure do not automatically coincide with the classical limit, as one might expect: for $\Lambda$ not much larger than unity, there is a clear discrepancy between the converged energies of the TB plasmons and the classical plasmons. Wherever the TB curves in Fig.~\ref{fig:constant_lambda} have become (almost) horizontal, the scale invariance that holds exactly for Dirac and classical plasmons has also emerged for TB quantum plasmons.

The bending of the curves for smaller widths illustrates the shortcomings of the scalability of the Dirac model: For it to hold exactly, we would need infinitely many bands in the band structure, but as the number of atoms in the full-width supercell decreases as $W$ is reduced (recall Fig.~\ref{fig:ribbongeometry}), we will get fewer bands instead and thus a deviation from the converged constant plasmon energy as obtained for wide ribbons.
In Fig.~\ref{fig:constant_lambda} we also see that the AC plasmon energies in general are closer to the classical predictions than the ZZ plasmon energies, as could also be extracted from Fig.~\ref{fig:contribs_and_widths}.

\begin{figure}[!htbp]
\centering
\includegraphics[width=3.4in]{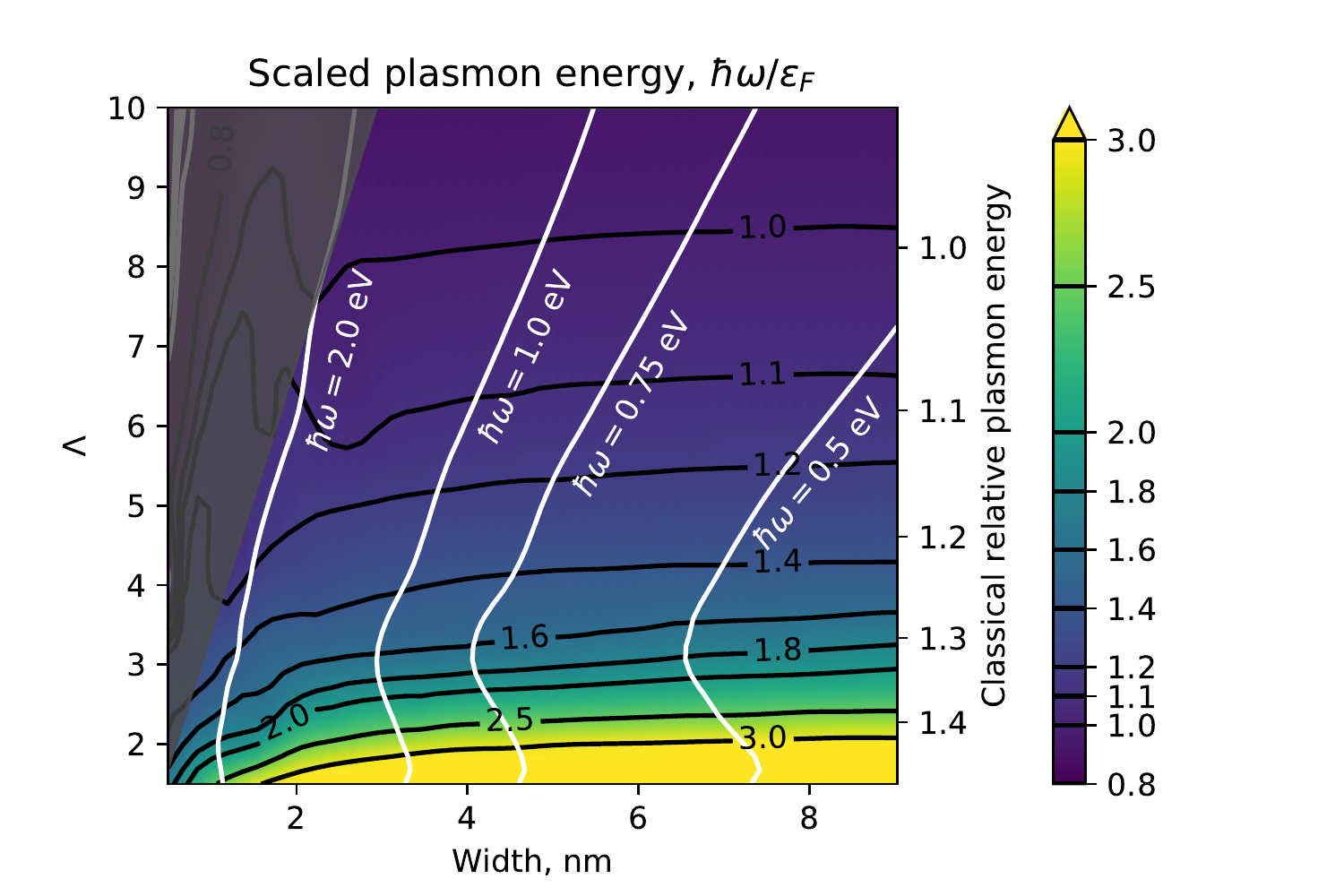}
\caption{(color online) The scaled plasmon energy as a function of the width and $\Lambda$ for zigzag ribbons. The white lines indicate paths with constant (absolute) plasmon energy. The gray area corresponds to structures with $\varepsilon_F>2\unit{eV}$ where we expect to be outside the linear regime of the bands. For larger widths the contours start to converge as expected from Dirac theory. The departure from horizontal lines is a signature of having only a finite number of electronic states.  
The right $y$-axis shows the classical results for comparison.}\label{fig:width_lambda}
\end{figure}

In Fig.~\ref{fig:width_lambda} we display the same data for zigzag ribbons as in Fig.~\ref{fig:constant_lambda} but in a complementary way, now as a function of width and $\Lambda$.  We obtain a surface plot of the scaled plasmon energy, where data points have been cubicly interpolated to get a smooth surface.
From this view, we also see the convergence of the scaled  plasmon energies that deviates significantly for the classical values displayed on the right $y$-axis.
For $\varepsilon_F>2.0\unit{eV}$ the Fermi energy is outside the linear part of the band structure and we find large deviations from the Dirac model as can be seen in the gray area of the plot.
By multiplying with the corresponding Fermi energy surface $\varepsilon_F = \hbar v_F\Lambda/W$ we calculate lines of constant plasmon energy (the white lines) thus conveniently providing a continuous range of options in parameter space to obtain a specific plasmon energy.
The bending of the white lines also reveals the offset from the classical behavior:
For large values of $\Lambda$ the system is well described when only including the classical intraband Drude term.
This results in a scaling of the scaled plasmon energy with $\Lambda^{-1/2}$ leading to the plasmon energy scaling as $\hbar\omega \propto \Lambda^{1/2}$.
The intraband divergence at $\Lambda = 0$ is quenched due to the screening of the interband transitions when the latter are included.
Looking at the data in Fig.~\ref{fig:width_lambda}, for a constant width, when going to smaller values of $\Lambda$, around $\Lambda_\mathrm{c}$ the plasmon energies start to increase again showing that the scaled plasmon energy must increase faster than as $\Lambda^{-1}$.

\section{Conclusions}\label{Sec:conclusions}

Using tight-binding calculations and inspired by Dirac theory, we identify multiple interesting effects in graphene nanoribbon plasmons:
As a first main result, for both armchair and zigzag ribbons an emerging non-classical scale-invariant behavior of the plasmon energies has been predicted and confirmed to exist also in TB calculations, even though the scale invariance does not hold strictly in the mathematical sense in TB. For ribbons wider than 5~nm, we illustrated in Fig.~\ref{fig:constant_lambda} that the scale invariance effectively holds for the energies considered, and better so for higher Fermi energy (which we kept smaller than 2.0 eV). The scale invariant horizontal curves that the TB calculations converge to generally have  non-classical limiting values. This is the realm where Dirac theory can be accurate, and only in the limit $\Lambda \equiv k_{F} W \gg 1$ do our TB plasmon energies agree nicely with those of classical plasmons. 

An experimental realization that comes close to the non-classical regime is the work in Ref.~\onlinecite{BrarHighlyNanoresonators} where the plasmonic mode of a 15 nm ribbon with $\epsilon_F\approx 0.4\unit{eV}$ has been measured.
This is still within the classical regime as $\Lambda \approx 9.3 > \Lambda_\mathrm{c}$, but lowering the Fermi energy to around $0.1\unit{eV}$ corresponding to $\Lambda \approx 2.3$ should reveal new quantum effects for both zigzag and armchair edge terminations.

As our second main result, we have related the energy of the bottom parabolic band at the $K$ points to the onset of the deviation from the classical model and calculated these energies analytically using the Dirac model. Here again, we find that the agreement between our heuristic analytical estimates and numerical calculations holds quite well and in a scale-invariant way, i.e. the analytical estimates describe the onset of non-classical plasmonics both for narrow ribbons with higher Fermi energies and wider ribbons with lower $\varepsilon_F$. Spectral differences between quantum and classical plasmons emerge slightly earlier for zigzag than for armchair ribbons (i.e for larger $\Lambda$, or already  for wider ribbons at equal Fermi energy). 

Third, for armchair nanoribbons we observe the disappearance of the plasmons at two different low values of the scaled ribbon width $\Lambda$, dependent on whether the ribbons are semiconducting or semi-metallic in their neutral state.
Fourth, for zigzag ribbons we have provided a convenient way of predicting absolute plasmon energies from the iso-frequency curves in Fig.~\ref{fig:width_lambda}.  
Finally, we  revealed how the edge states of nanoribbons contribute with a significant blueshift of plasmon energies, in contrast to reported redshifts for other graphene nanostructures.

\section*{Acknowledgments}
We would like to thank Thomas Christensen, Johan R. Maack and P. André D. Gon\c{c}alves for stimulating discussions.
This work was supported by the
Danish Council for Independent Research--Natural Sciences (Project 1323-00087). 
The Center for Nanostructured Graphene is sponsored by the Danish National Research Foundation (Project No. DNRF103).
N.~A.~M. is a VILLUM Investigator supported by VILLUM FONDEN (grant No. 16498).

\appendix

\section{Identifying edge states in tight-binding}\label{app:edgyness}

In the Dirac model the edge states of graphene zigzag ribbons are readily found as solutions that decay exponentially fast from the edge of the structure in contrast to the bulk-like modes that behave more like standing waves.
As the tight-binding model is solved numerically by diagonalizing the Hamiltonian we do not get this distinction for free, but need to analyze the resulting states subsequently in order to classify them properly.
To give an overview of where in the band diagrams calculated in TB we find these edge states, we introduce an operational definition of \emph{edginess} as
\begin{align}\label{Eq:edginess}
\lambda_n(k) = \frac{\sum_{l\in\Omega}\abs{\psi_{nl}(k)}^2 - \sum_{l\notin\Omega}\abs{\psi_{nl}(k)}^2}{\sum_l\abs{\psi_{nl}(k)}^2},
\end{align}
where $l$ refers to the atomic sites.
In other words, the edginess $\lambda_n$ of the $n$'th state is found by the amount of the wavefunction localized on the edge of the ribbon, $\Omega$, subtracted with the weight in the middle of the ribbon.
In our case we define $\Omega$ as the outermost quarters of the atoms on either side of the ribbon.
Using this definition, an edge mode will have $\lambda\simeq1$ while eigenstates located entirely in the center of the ribbon will have $\lambda = -1$.

\section{Coulomb interaction in real space}\label{app:coulomb}

At large distances, the Coulomb interaction between two sites will be predominantly point-like and thus scale as their inverse distance. This long-range $r^{-1}$ behavior makes it practically impossible to calculate the correct Coulomb interaction term in real space. Fortunately this is not necessary either, as we ultimately are interested in the dielectric function. 
Instead, as we require charge neutrality, we utilize that $\sum_i \chi^0_{ij} = 0$ to calculate a modified interaction\cite{Thongrattanasiri2012QuantumPlasmons}
\begin{align*}
\tilde V_{ij} = \sum_{n,j} \pa{ V_{i0,jn} - \abs{nb}^{-1} },
\end{align*}
which fulfills $V\chi^0 = \tilde V\chi^0$ and  falls off more quickly with distance than $V$.
We have used the notation $V_{i0,jn}$ to mean the interaction between the $i$'th site in the 0'th supercell and the $j$'th site in supercell $n$.

In the short-distance limit, keeping the assumption of point-like interactions would lead to a diverging Coulomb term for for the distance going to zero. Instead, for sites close to each other and, ultimately, for a site interacting with itself, the spatial extent of the $p_z$-orbitals should be taken into account. This has been done by Ref.~\onlinecite{Thongrattanasiri2012QuantumPlasmons} and we adopt the same approach for all distances (with data acquired through private correspondence between our groups), whereby the Coulomb term no longer diverges for vanishing distances.

\bibliographystyle{wubssty}
\bibliography{Mendeley}

\end{document}